\documentclass[12pt,preprint]{aastex}

\newcommand{\Msun}{M$_{\odot}$}
\newcommand{\kms}{km s$^{-1}$}
\newcommand{\etal}{{\it et al.}}

\begin{document}

\title{Effects of Uniform and Differential Rotation on Stellar Pulsations}
\author{C.C. Lovekin\altaffilmark{1}}
\affil{Institute for Computational Astrophysics, Department of Astronomy and Physics, Saint Mary's University}
\email{clovekin@ap.smu.ca}
\author{R.G. Deupree}
\affil{Institute for Computational Astrophysics, Department of Astronomy and Physics, Saint Mary's University}
\author{M.J. Clement}
\affil{Department of Astronomy and Astrophysics, University of Toronto}
\altaffiltext{1}{Current address:  LESIA, Observatoire de Paris, Meudon}

\begin{abstract}
We have investigated the effects of uniform rotation and a specific model for differential rotation on the pulsation frequencies of 10 \Msun\ stellar models.  Uniform rotation decreases the frequencies for all modes.  Differential rotation does not appear to have a significant effect on the frequencies, except for the most extreme differentially rotating models.  In all cases, the large and small separations show the effects of rotation at lower velocities than do the individual frequencies.  Unfortunately, to a certain extent, differential rotation mimics the effects of more rapid rotation, and only the presence of some specific observed frequencies with well identified modes will be able to uniquely constrain the internal rotation of pulsating stars.
\end{abstract}

\section{Introduction}

Observationally detected stellar pulsation frequencies can be used to place constraints on stellar models, giving us an improved understanding of the interior structure and evolution of stars.  The most successful application has been the Sun, where the large number of observed modes have placed strict constraints on parameters such as the helium abundance (Y) \citep{basu04,antia06}, the depth of the convection zone \citep{cdgt89,cdgt91}, and the interior angular momentum distribution \citep{darwich,couvidat}.  Observations of pulsation frequencies of other stars continue to improve, particularly through dedicated satellites such as WIRE \citep{wire}, MOST \citep{most}, CoRoT \citep{corot} and Kepler \citep{kepler} as well as ground-based networks such as STEPHI \citep{steph} and WET \citep{wet}.  These improved observations, giving us long term coverage and improved accuracy, are the first steps in enabling other stars to be constrained in a similar manner to the Sun.  Asteroseismology then, has the potential to answer a number of questions about the interior structure of stars throughout the HR diagram.  

One aspect of stellar structure which could be explored using asteroseismology is the internal rotation rate.  It is theoretically possible for stars to rotate with angular velocity increasing or decreasing with distance from the rotation axis, and there is some evidence that the latter may be true in massive main sequence stars, at least at the surface \citep{stoeckley}. A third possibility is uniform rotation.  It has been argued that uniform rotation is unrealistic based on observations of the Praesepe and Hyades clusters \citep{smith71}.  Of course, other, less well structured rotation laws are possible.  However, there is little evidence in support of a specific rotation law, and the large uncertainties prevent any of the possibilities from being ruled out.  We note in passing that the solar rotation rate in the convection zone is primarily dependent on latitude \citep{schou98,thompson03} and thus cannot be described by a conservative rotation law. 

Recently, interferometric observations of Achernar \citep{dom03} found that this star is far more oblate than is possible for a uniformly rotating star. This is true because uniformly rotating stars reach critical rotation before they have sufficient angular momentum to produce such an oblate object. However, \citet{jack05} noted that models in which the rotation rate increases inward from the surface can produce the oblateness observed for Achernar and still match the observed $v$sin$i$. While further study has proposed that the oblateness may be due to a circumstellar envelope \citep{carcofi}, the original conclusion does raise the interesting question as to whether stars with rotation laws required to produce such an oblate shape exist, and if so, whether these rotation laws could be identified by possible pulsation modes. We investigate this possibility in this paper. 

Differential rotation with the rotation rate increasing inwards, as is considered in this paper, will have an impact on the deep interior structure of the star, provided the differential rotation is large enough.  Rapid rotation in the outer layers of a star has little to no effect on the gravitational potential and core structure, as the envelope contains a relatively small fraction of the stellar mass.  In fact, many early attempts to model rotating stars assumed that the mass in the envelope was negligible and that the gravitational potential in this region could be modelled using a Roche potential \citep{sackmann70}.  However, it was recognized early on that this assumption was not always valid.  Efforts to model a wider variety of rotating stars were made through the implementation of the self-consistent field (SCF) method \citep{ostriker}, or through direct, 2D finite difference solutions to Poisson's equation \citep{clement74,clement78,clement79}.  These methods allowed stars to be modelled with differential rotation, at least under certain circumstances.  Concentrating angular momentum in the center, unlike uniform rotation, can produce enough distortion to affect the core, and consequently the evolution of the star.  Only in this case can we produce a model with interior properties significantly different from the uniformly rotating model \citep{sackmann70}.  Even restricting ourselves to this type of differential rotation does not narrow the possibilities significantly.  The rotation could be shellular, as proposed by \citet{zahn92}, or cylindrical (conservative rotation laws).  In this paper, we have focused on conservative rotation laws, either with uniform rotation or with the rotation rate increasing towards the center of the star.  Further discussion of our models can be found in \S \ref{models}.  In \S \ref{relfreq} we consider the eigenfrequencies of rotating models as well as the large and small separations in \S  \ref{large} and \S \ref{small} respectively.  Our conclusions are summarized in \S \ref{conclusion}.

\section{Numerical Models}
\label{models}

The stellar models are computed using the 2D stellar structure code ROTORC \citep{bob90,bob95}.  The code uses the OPAL opacities \citep{opalk} and equation of state \citep{opaleos}. Here we consider only 10 \Msun\ ZAMS models with X=0.7, Z=0.02. These models solve the conservation equations of mass, momentum, energy, and hydrogen abundance along with Poisson’s equation for the gravitational potential on a two dimensional finite difference grid with the fractional surface equatorial radius and the colatitude as the independent variables. The surface equatorial radius is determined by requiring that the integral of the density over the volume of the model equals the stellar mass. The ZAMS models are taken to be time independent and static, except for the imposed rotation law, so that the mass, azimuthal momentum, and hydrogen composition conservation equations drop out. 

The only change required in the stellar evolution code for nonuniform cylindrical rotation laws is the addition of an extra term in the total potential (e.g., Tassoul 2000):
\begin{eqnarray}
\Psi &=& \Phi - \int_0^{\varpi}\Omega^2(\varpi')\varpi'd\varpi' \\ \nonumber
 & = & \Phi - \frac{\Omega^2\varpi^2}{2} + \int_0^{\varpi}\varpi'^2\Omega(\varpi')\frac{d\Omega(\varpi')}{d\varpi'}d\varpi'
\end{eqnarray}
where $\Omega$ is the rotation velocity (in radians per second) and $\varpi$ is the distance from the rotation axis ($xsin\theta$, where $x$ is the fractional surface equatorial radius and $\theta$ is the colatitude).  The extra term is the last term on the right hand side of the equation. Performing the integral is straightforward for an analytically imposed rotation rate distribution in the ZAMS models. The total potential is used only to determine the surface location at each latitude by taking the surface to be an equipotential.  Although defining a total potential requires a conservative rotation law, this is the only way in which a conservative rotation law is utilized in the stellar structure code.

\begin{figure}
\plotone{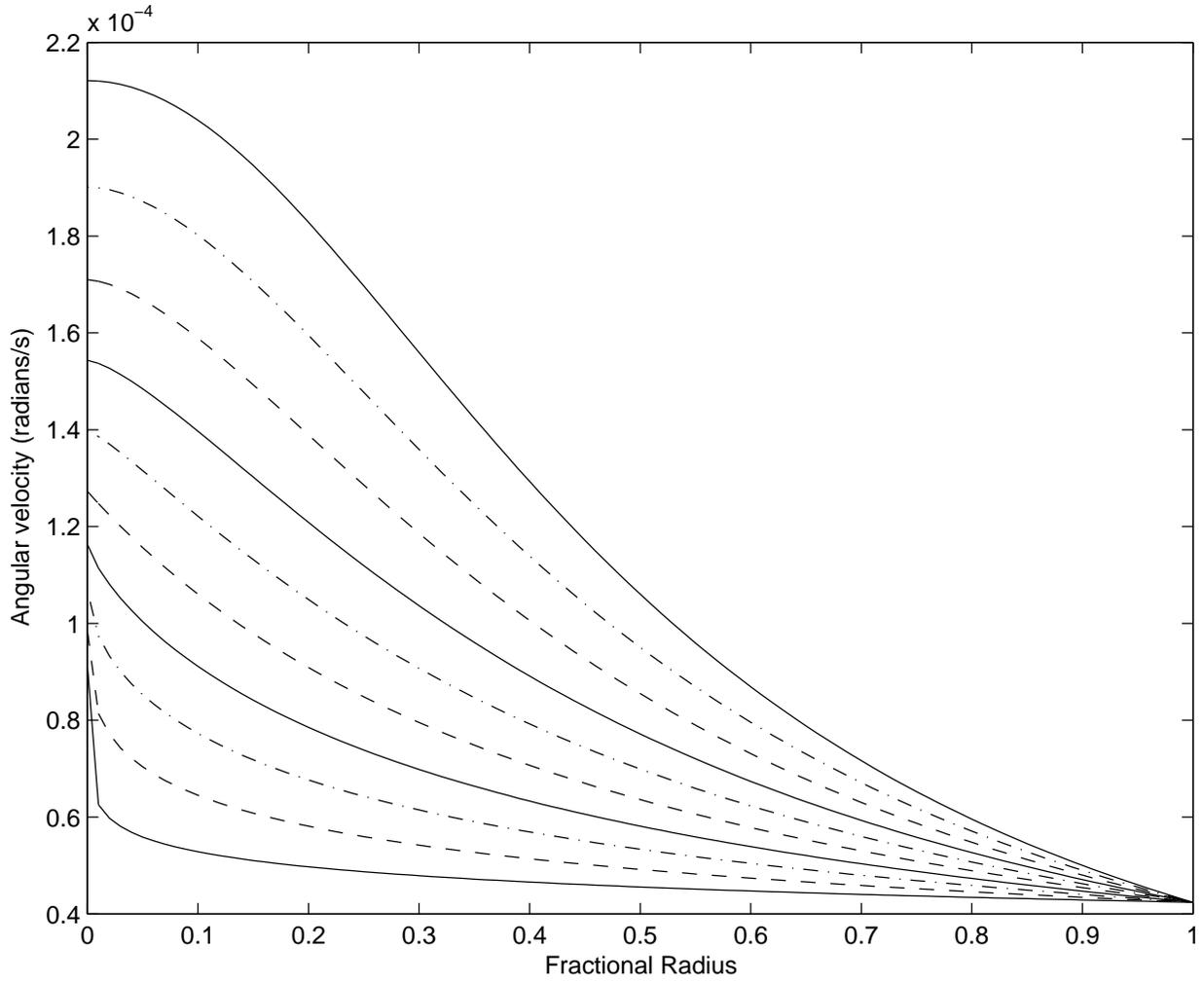}
\caption{\label{fig:omega}Rotation law used in differentially rotating models (Eqn. \ref{eqn:omega}).  Curves show from bottom to top the rotation law for $\beta$ = 0.2, 0.4, 0.6, 0.8, 1.0, 1.2, 1.4, 1.6, 1.8 and 2.0 for a model with surface equatorial velocity of 120 \kms.}
\end{figure}

We have constructed uniformly rotating ZAMS models with rotation velocities between 0 and 360 \kms, with an approximate spacing of 30 \kms. We have also computed a number of differentially rotating models at two values of the surface equatorial rotation velocity, 120 and 240 \kms.\@ The differential rotation law is as given by \citet{jack05}:
\begin{equation}
\label{eqn:omega}
\Omega(\varpi) = \frac{\Omega_o}{1+(a\varpi)^{\beta}}
\end{equation}
where $\beta$ is a parameter ranging from 0 (uniform rotation) to 2, the maximum allowed for stability. The parameters $a$ and $\Omega_o$ are used to impose the desired surface equatorial velocity and shape of the rotation law at small distances from the rotation axis.  We have arbitrarily chosen $a = 2$.\@  Figure \ref{fig:omega} shows the rotation rate as a function of distance perpendicular to the rotation axis for a surface equatorial rotation velocity of 120 \kms\ and a surface equatorial radius for a uniformly rotating model at that speed.  Increasing $\beta$ increases the rotation rate close to the the rotation axis, including in the core of the star.  Increasing angular momentum increases structural changes, and thus the structural changes increase with increasing $\beta$.  It is expected that increasing the rotation rate through increasing $\beta$ may in some ways mimic more rapid uniform rotation.

\begin{figure}
\plotone{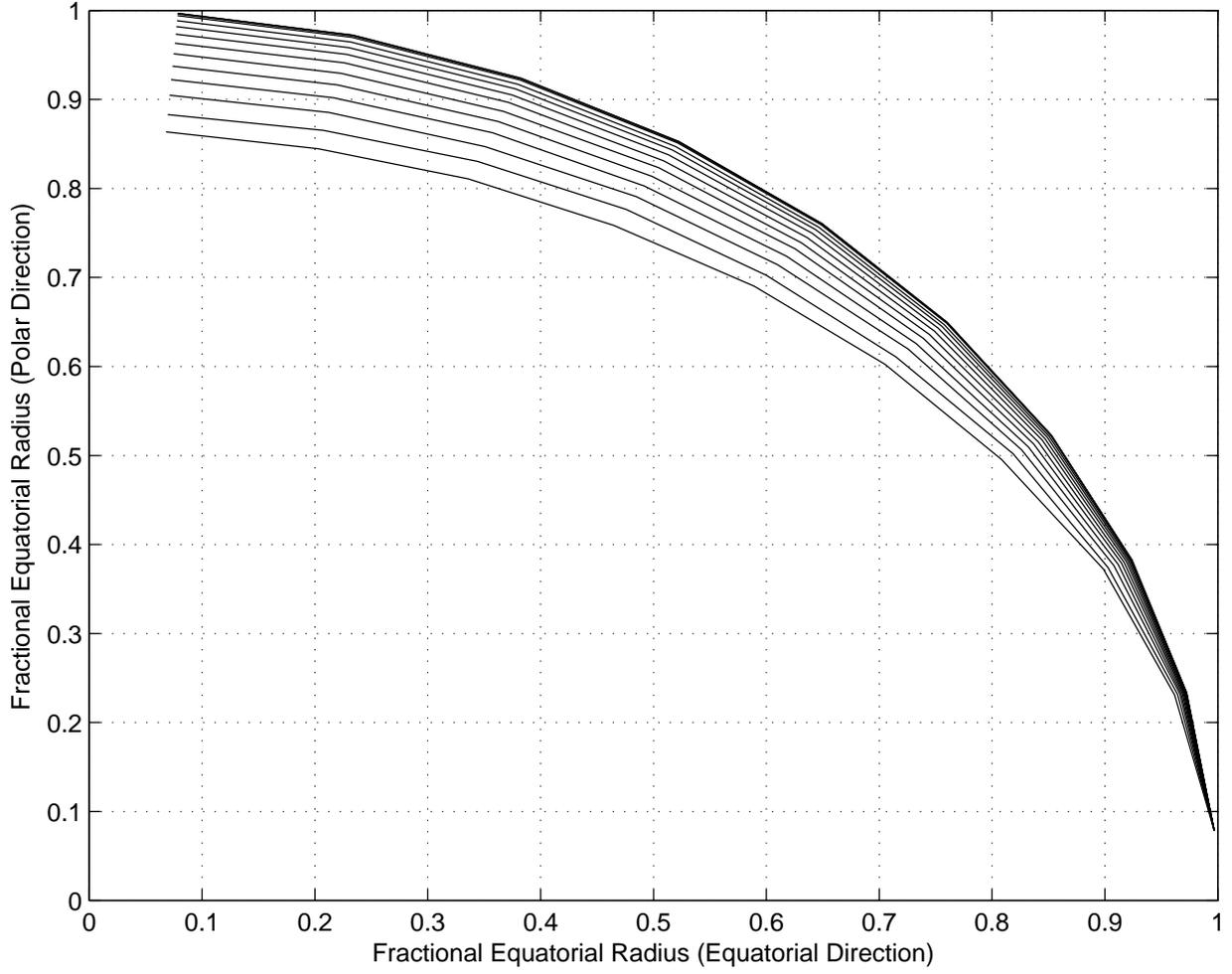}
\caption{\label{fig:shapeuniform}Surface shape for uniformly rotating models.  The polar radius decreases relative to the equatorial radius as rotation increases from 0 \kms\ to 360 \kms.}
\end{figure}

\begin{figure}
\plotone{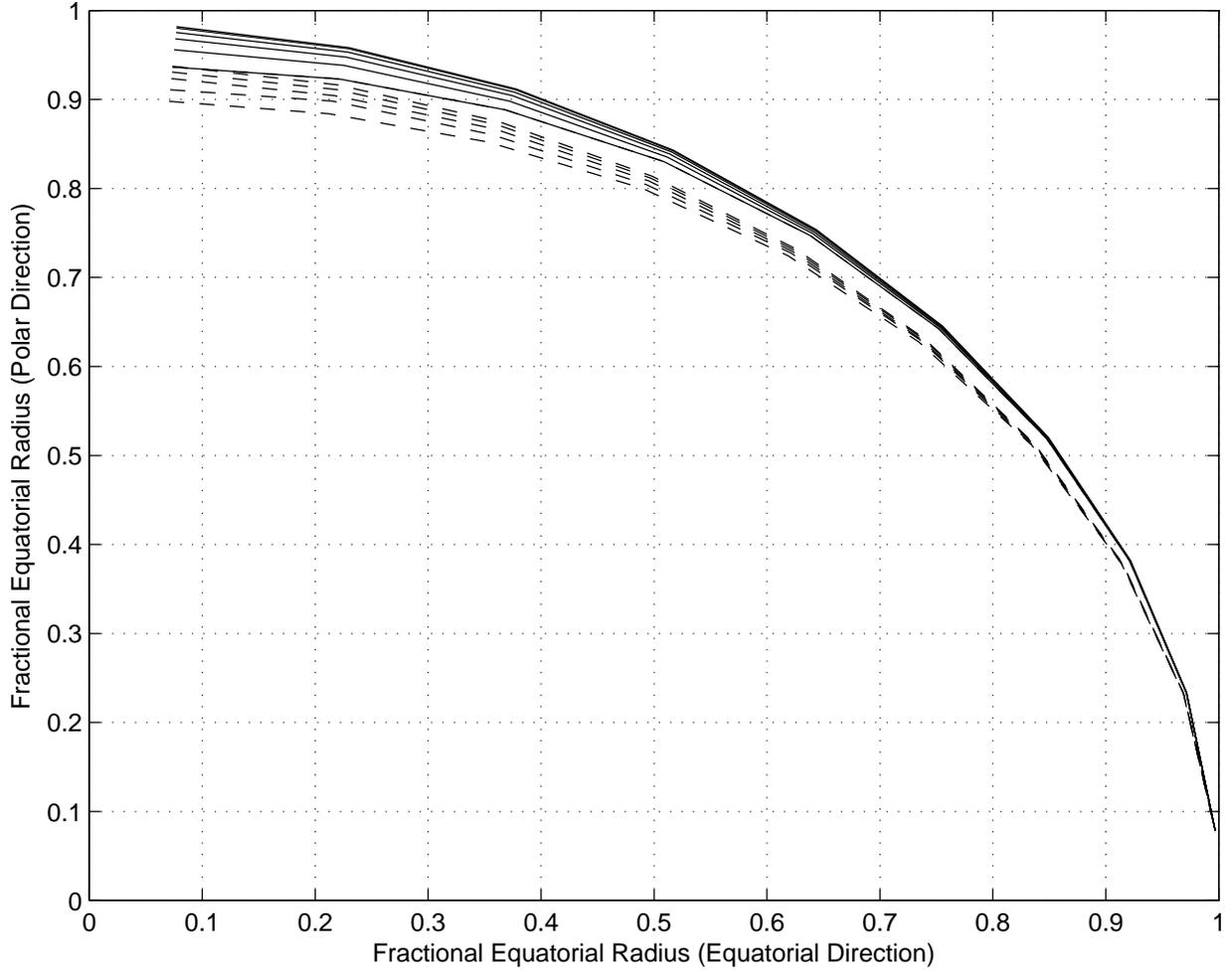}
\caption{\label{fig:shapediff}Surface shape for differentially rotating models at 120 \kms\ (solid) and 240 \kms\ (dashed).  As the rotation rate close to the rotation axis increases (increasing $\beta$), the polar radius decreases relative to the uniformly rotating case.}
\end{figure}

One major result produced by significant rotation is an appreciable distortion of the surface of the model. We present the surface shape for a set of uniformly rotating models with surface equatorial velocities ranging from 0 to 360 \kms\ in Figure \ref{fig:shapeuniform}.\@ For each model the equatorial radius is taken to be unity. The ratio between the polar and equatorial radius decreases with increasing rotation, as the polar radius decreases slightly while the equatorial radius increases considerably. Differential rotation in which the rotation rate increases with decreasing distance from the rotation axis amplifies this effect. We present the surface shape for the differentially rotating models in Figure \ref{fig:shapediff}. The solid curves are for a surface equatorial velocity of 120 \kms, while the dashed curves denote a surface equatorial velocity of 240 \kms. As the parameter $\beta$ in Equation \ref{eqn:omega} increases, the fractional polar radius decreases. The change in fractional radius with $\beta$ is greatest at the pole and decreases towards the equator.  Note that the fractional polar radius for a model rotating with a surface equatorial velocity of 120 \kms\ and a value of $\beta$ of 1.8 has nearly the same fractional polar radius as a model uniformly rotating at 240 \kms.      

We have increased the radial resolution of the static models by more than a factor of two over that used by Lovekin and Deupree (2008). The intent is to reduce the scatter and uncertainty in the pulsation mode calculations, particularly for the large and small separations. By and large this has been successful. 

The determination of the pulsational properties of these models is made using the linear adiabatic pulsation code developed by \citet{NRO}. We restrict our attention to input models with conservative rotation laws so that we can write the effective gravity ($\vec{g}$) as the derivative of the total potential. The input models are axisymmetric spheroids, which allows us to assume a $e^{i[\omega t + m \phi]}$ time and azimuthal dependence.  The equations to be linearized are the three components of the momentum conservation equation, the mass conservation equation, the adiabatic relation between the density ($\rho$) and the pressure (P), and Poisson's equation.  The dependent variables are the three components of the linearized displacements ($\vec{\xi}$), the linearized Eulerian displacements of the density ($\delta\rho$) pressure ($\delta P$), and the gravitational potential ($\delta\phi$).  We start with the perturbed momentum equation:
\begin{equation}
\delta\left(\frac{d\vec{v}}{dt}\right) = \delta\left(\frac{\partial \vec{v}}{\partial t}+\left[\vec{v}\cdot\triangledown\right]\vec{v}\right) = \triangledown(\delta\phi)+\left(\frac{\delta\rho}{\rho^2}\right)\triangledown P - \frac{1}{\rho}\triangledown\delta P
\end{equation}
where $\vec{v}$ is the velocity, which includes both the pulsational perturbations and the velocity distribution of the unperturbed model  Here we assume this later velocity results exclusively from rotation, and is given by
\begin{equation}
\vec{v} = v\hat{\phi} = \Omega r sin(\theta)\hat{\phi}.
\end{equation}
Here $\Omega$ is the rotation rate, and our assumption of a conservative rotation law requires it to be a function only of the distance from the rotation axis, $\varpi = rsin(\theta)$.  We can write an expression for the Eulerian perturbation of the velocity in terms of the displacements because the Lagrangian time derivative of the displacement is the Lagrangian velocity perturbation.  Thus,
\begin{equation}
\delta\vec{v} = \frac{\partial\vec{\xi}}{\partial t} + (\vec{v}\cdot\triangledown)\vec{\xi}-(\vec{\xi}\cdot\triangledown)\vec{v}.
\end{equation}
Substituting this expression into the above perturbed momentum equation and proceeding along the same lines as \citet{NRO} leads to the following expression for the three individual components of the momentum equation:
\begin{equation}
\sigma^2\xi_r + 2i\sigma\Omega\xi_{\phi}sin\theta - \varpi sin\theta\left[\xi_r\frac{\partial\Omega^2}{\partial r}+\frac{\xi_{\theta}}{r}\frac{\partial\Omega^2}{\partial\theta}\right]-\frac{\partial\delta p}{\partial r} = -Ag_r(\triangledown\cdot\vec{\xi})
\end{equation}
\begin{equation}
\sigma^2\xi_{\theta} + 2i\sigma\Omega\xi_{\phi}cos\theta-\varpi cos\theta\left[\xi_r\frac{\partial\Omega^2}{\partial r} + \frac{\xi_{\theta}}{r}\frac{\partial\Omega^2}{\partial\theta}\right]-\frac{1}{r}\frac{\partial\delta p}{\partial \theta} = -Ag_{\theta}(\triangledown\cdot\vec{\xi})
\end{equation}
\begin{equation}
\sigma^2\xi_{\phi} - 2i\sigma\Omega(\xi_r sin\theta + \xi_{\theta}cos\theta)-\frac{im}{rsin\theta}\delta p = 0,
\end{equation}
where $\sigma = \omega + m\Omega$, $\delta p = \delta P/\rho - \delta \phi$, $A = c^2dln\rho/d\Psi - 1$ and $c^2 = \Gamma_1P/\rho$.  The quantity $c$ is the adiabatic sound speed.  The other equations required are as listed in \citet{NRO}:  mass conservation, the adiabatic relationship between the density and pressure perturbations, and Poisson's equation for the perturbed gravitational potential:
\begin{equation}
\delta\rho = -\triangledown\cdot\rho\xi
\end{equation}
\begin{equation}
\delta P = -\Gamma_1 P\triangledown\cdot\xi - \xi\cdot\triangledown P
\end{equation}
\begin{equation}
\triangledown^2\delta\phi = -4\pi G\delta\rho.
\end{equation}

Differential rotation has introduced two new terms into each of the radial and latitudinal momentum equations.  The other change is that $\sigma$, the eigenfrequency in the local rotating frame, is no longer constant except for axisymmetric modes.  This does not alter the solution algorithm, however, because Clement's (1998) approach was to select $\sigma$ (which now becomes selecting $\omega$ and calculating $\sigma$ locally as needed),
solve all the equations and see if a discriminant was satisfied.  Neither the added terms nor the nature of the eigenvalue requires any further modification to the approach, and the derivation of the final equations proceeds as described by \citet{NRO}.  

Using these equations we can calculate the pulsation properties of the stellar models on a 2D finite difference grid.  This is done through a change of variables, factoring out the behavior of $\xi_r$, $\xi_\theta$, $\delta p$ and $\delta\phi$ near the boundaries to eliminate singularities.  The coefficients of these equations can be put in a band-diagonal matrix and solved.  {\tt NRO} can include up to nine angular zones in the eigenfunction solution.  This gives the solution at N angles, where N is the number of angular zones, which can subsequently be decomposed into the contributions of individual spherical harmonics through the use of Fourier transforms.  Throughout this paper, we have used N = 6.  Based on the calculations of \citet{me}, six spherical harmonics is sufficient to accurately calculate the eigenfrequencies for the most rapidly rotating models discussed here.  Indeed, we have performed a few test calculations with N = 8 and have found that the effect on the frequencies is small, typically a few hundredths of a percent.

As discussed in \citet{me}, {\tt NRO}, combined with stellar structure models from {\tt ROTORC}, allows us to calculate the pulsation frequencies for rotating stars without making any {\it a priori} assumptions about the structure, except that the rotation law is conservative for {\tt NRO}.  For further discussion of the method of solution used by NRO, refer to \citet{NRO} or \citet{me}.

With spherical stellar models, the radial and angular components of the perturbations separate, and the angular part can be expressed as a spherical harmonic with specific values of the quantum numbers, $l$ and $m$.  

For rotating stars, the eigenfunction solution is not a single spherical harmonic, and $l$ is not a valid quantum number.  Indeed, {\tt NRO} uses $l$ only to specify the parity of the mode being calculated, and includes the first $k$ even or odd spherical harmonics, where $k$ is the number of angles included.  We identify modes using a quantum number $l_o$, which is the $l$ of the mode in the non-rotating model to which a given mode can be traced.  For spheroids, $m$ remains a valid quantum number.  As in \citet{me}, we restrict ourselves to axisymmetric modes ($m=0$) and modes with small radial quantum number ($n$).



\section{Relative frequencies}
\label{relfreq}

In this paper we consider low order axisymmetric modes for $l_o$ = 0, 1, 2 and 3.  These modes are expected to have the highest amplitudes and the smallest cancellation effects across the visible surface of the star, and are hence expected to be the most easily visible.  Our structural models cover velocities from 0 to 360 \kms\ and for two velocities, 120 and 240 \kms, we have calculated differentially rotating models with $\beta$ varying from 0 to 2.0.  Tracing the individual modes becomes very difficult above rotation velocities of 360 \kms\ and for some higher values of $\beta$, and this represents a practical limit to our study. Although the frequencies can be calculated at these velocities, the resulting eigenfunctions are a mix of six spherical harmonics, and no single harmonic dominates.  As it is very difficult to reliably assign a value of $l_o$ to these modes, we exclude them from our analysis.  It is probably feasible to trace the modes accurately, but this could require an extremely fine rotational velocity grid (1-5 \kms).  We decided not to pursue this for this exploratory work.  For differentially rotating models, the limits are $\beta$ = 1.8 for the 120 \kms\ model and $\beta$ = 1.0 for the 240 \kms\ model.  Based on the curves shown in Figure \ref{fig:omega}, it appears that the limit is related to the angular velocity near the rotation axis.  The curve representing $\beta$ = 1.0 has approximately half the value at the center of the $\beta$ = 2.0 curve.  Therefore, if we double the velocities everywhere, the limiting $\beta$ should move from 1.8 at 120 \kms\ to 1.0 at 240 \kms, which corresponds to approximately the same angular velocity near the rotation axis.

\subsection{Uniform Rotation}
\label{sec:unifreq}

\begin{figure}
\plotone{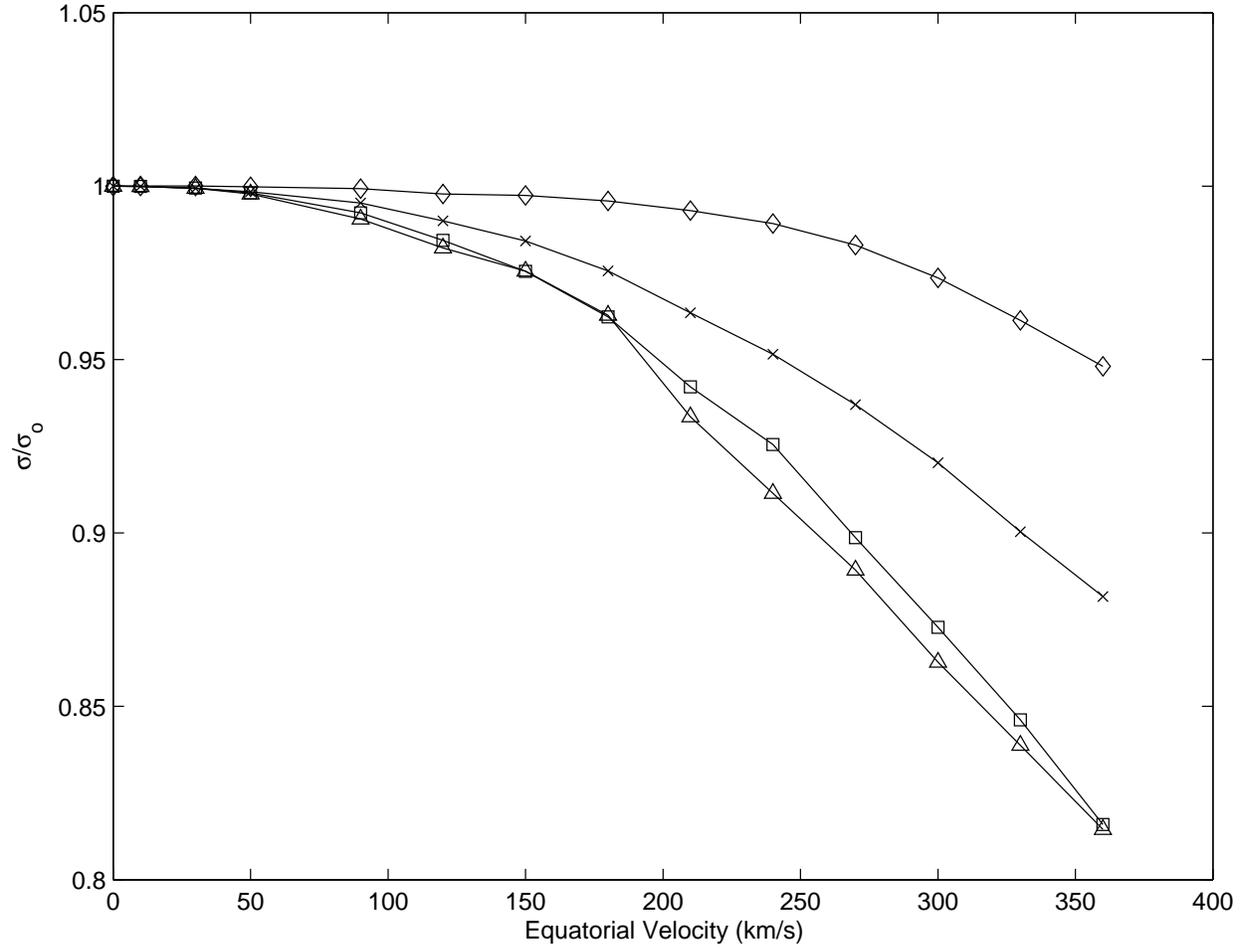}
\caption{\label{fig:l0}The first four harmonics of the $l_o$ = 2 mode for a uniformly rotating 10 \Msun\ model as a function of the rotation rate.  The four curves represent the frequencies for the $f$ (diamond), $p_1$ (X), $p_2$ (square) and $p_3$ (triangle) modes.}
\end{figure}

The trends produced by tracing a given mode through increases in rotation velocity are illustrated in Figure \ref{fig:l0} for the $l_o$ = 2 mode, which shows the eigenfrequencies normalized by the non-rotating frequency for each mode.  Overall, the trends we find for frequency agree with those calculated by previous work \citep{lignieres}.  These authors find that the frequencies decrease as one increases the rotation rate, with higher frequency modes decreasing more than lower modes.  As discussed in \citet{me}, our results at low to moderate rotation rates are also consistent with the frequency trends predicted by second order perturbation theory (see for example, Saio 1981).

We have increased the radial resolution of the outer 30\% of the radius of the static models by more than a factor of two over that used by \citet{me}. This produces a radial zoning finer than that currently allowed by the pulsation code, so further increases in radial resolution in the 2D structure models will only be effective if the pulsation code is modified to allow more radial zones. The intent of the modified zoning is to reduce the scatter and uncertainty in the mode calculations evident in \cite{me}. Figure \ref{fig:l0} shows that a reasonable estimate of our accuracy for the frequencies is a very few tenths of a percent, although there are still a few frequencies, most commonly but not exclusively for the higher radial orders and higher rotation rates, which do not fit within this limit. One might expect the radial resolution near the rotation axis and at mid latitudes to be less than for lower rotation rates because the fractional radius at these latitudes compared to the equator is lower. The accuracy of the small separation appears well within a $\mu$Hz, while the large separation does show variations on the order of one $\mu$Hz, particularly at higher rotation velocities and for higher radial order modes. This is compatible with the notion that the radial resolution near the surface could continue to benefit from refinement. However, these uncertainties do not disguise trends in the results, even in the large separation, with respect to rotation rate or the rotation law, and we consider these trends significant.

\begin{figure}
\plotone{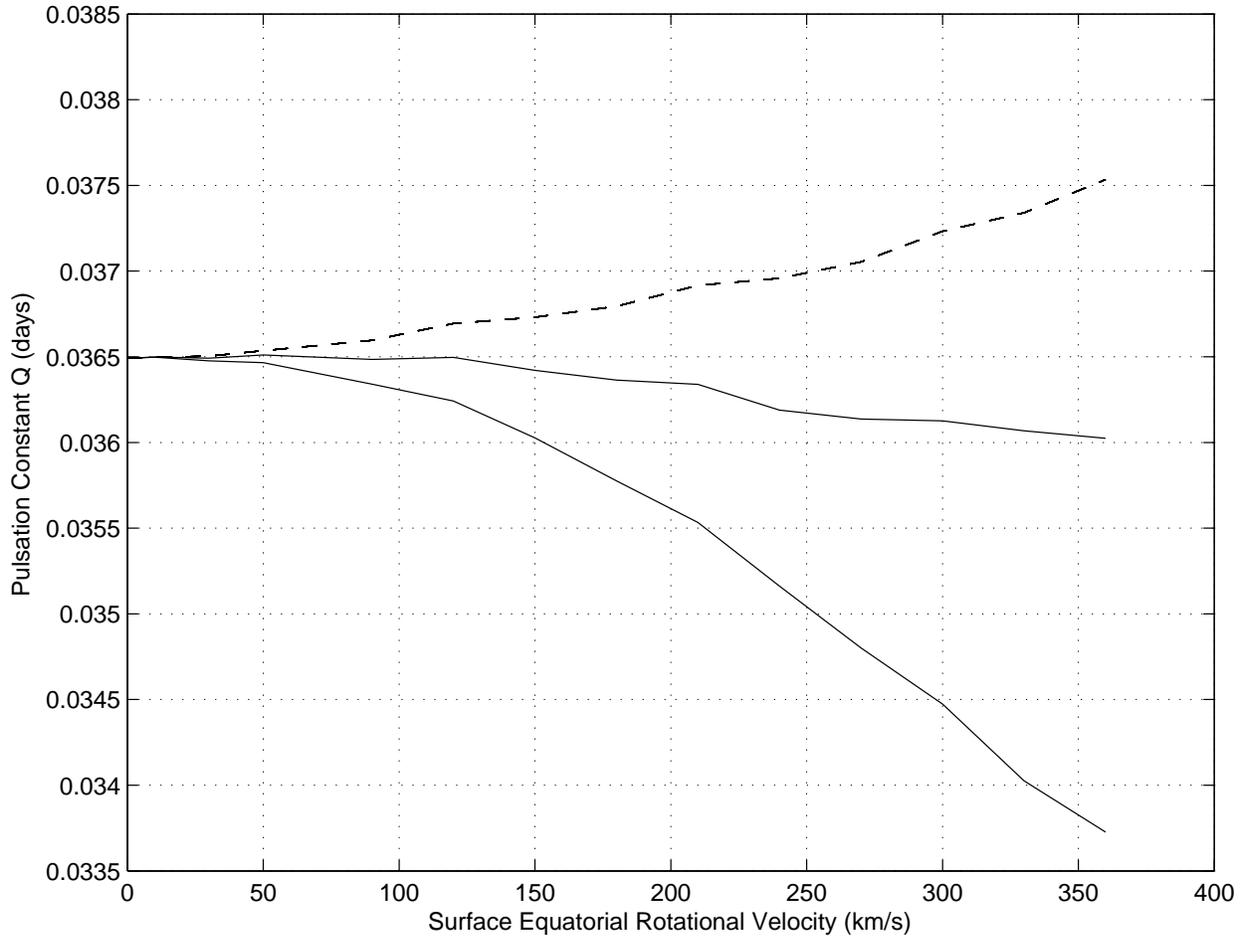}
\caption{\label{fig:q}Plot of the pulsation constant for the radius of a sphere with the same volume as the rotating model (bottom) and the radius at colatitude of 40$^{\circ}$ (middle).  Also shown is the pulsation constant assuming there is no change to the mean density as the rotation rate increases (dashed curve).}
\end{figure}

One interesting line of inquiry is whether there is some analog to the period - mean density relation which allows interpolation of eigenfrequencies as functions of models and rotation rates.  Specifically, we have examined if there is a physically meaningful radius which can be used in the period - mean density relation
\begin{equation}
\label{eqn:q}
Q = P \sqrt{\frac{M}{R^3}}
\end{equation}
(where M and R are in solar units, and P is the period) that would allow Q to be approximately constant as a function of the rotation rate.  The comparatively small changes in the eigenfrequencies shown in Figure \ref{fig:l0} suggest that the surface equatorial radius, with its fairly rapid increase as a function of rotation, will not keep Q approximately constant, and it does not.  The same is true for an average radius, defined as either a straight average or the effective radius required to contain the total volume of the model.  The polar radius would be more promising because it only slowly varies with the rotation rate, but it actually decreases slightly as the rotation rate increases.  This is the wrong direction to keep Q constant because the frequency decreases as well.  Because the polar radius decreases slightly and the equatorial radius increases appreciably with increasing rotation, one might guess there would be some latitude at which the radius increases at a rate that nearly offsets the rate of period increase.  This is true, and occurs at a colatitude of 40 degrees.  It is not obvious that this has any physical significance because it is difficult to associate any specific meaning to the radius at this latitude.  We present the pulsation constant for two definitions of an effective radius in Figure \ref{fig:q}. One way is to use the radius of a sphere with the same volume as the model.  The other uses the radius at a colatitude of 40$^{\circ}$.  For comparison we show a ``pulsation constant'' that would exist if we used the mean density of the nonrotating model.  We include this to give an idea of the size of the effect.  Interestingly, the variation in the pulsation constant is significantly larger allowing the mean density to be determined by the total volume of the rotating model (the mass is the same for all models) than it is when the mean density is assumed to be that of the nonrotating model.

The frequencies can also be changed by the mass or evolutionary state of the star, producing trends that could potentially be confused with rotational effects.  We wish to determine how closely the frequencies of a rotating model can be mimicked by a non-rotating model.  First  we calculated the Q values for each model in the 10 and 12 \Msun\ non-rotating models.  For each mode, we then took the mean of  the Q of the two models.  We used this average Q for the radial fundamental mode and the frequency of the radial fundamental mode for the model rotating at 150 \kms\ to calculate a mean density.  This corresponds to the mean density of a non-rotating model of unknown mass and radius pulsating in the radial fundamental mode with the same frequency as the 10 \Msun\ model rotating at 150 \kms.  The mean density found this way and the average Q's for the other $l_o$'s can be used to predict the other frequencies of this presumed non-rotating model.  When these frequencies are compared to the calculated frequencies for the rotating model, the differences are significant.  Using Q to calculate the frequencies in this way forces the radial fundamental mode to have the same frequency, so the differences between frequencies should be solely a result of rotation.  The frequencies predicted for the $l_o$ = 0 and 2 modes are larger by 1-5\%, with the differences increasing for higher order modes.  At the same time, the frequencies predicted for the $l_o$ = 1 and 3 modes are smaller, by as much as 15 \% for the $l_o$ =1 $p_1$ mode.  As the radial order increases, the differences between the rotating model and the non-rotating calculation decreases for the $l_o$ = 1 modes, but increases for the $l_o$ = 3 modes. The size and direction of these trends implies that the pulsation spectrum of a rotating model is unlikely to be confused with the pulsation spectrum of a more massive non-rotating model.  It also suggests that rotation must be included in the calculations if observations indicate it might be present even at this moderate amount.

We have also evolved a single non-rotating model, and compared the ZAMS model with one part way through the main sequence evolution (X$_c$ = 0.47).  In this case, the frequencies decreased sufficiently, even for a model with a large remaining core hydrogen fraction, that confusion seems unlikely.

\subsection{Differential Rotation}

\begin{figure}
\plotone{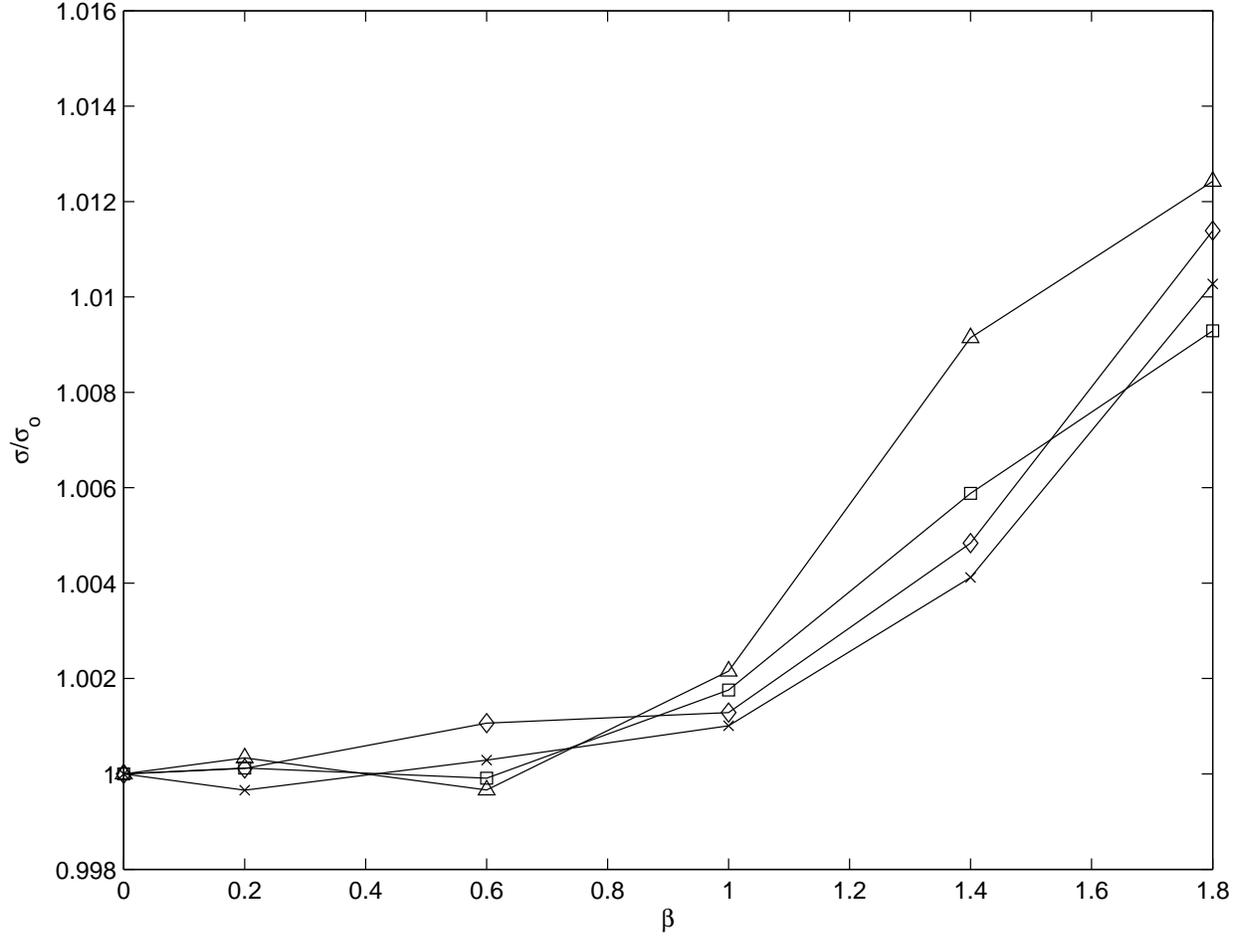}
\caption{\label{fig:120fund}The fundamental and first three overtones of the $l_o$ = 0  mode for a model rotating at 120 \kms as a function of the differential rotation parameter $\beta$ (see Equation \ref{eqn:omega}).  The four curves represent the frequencies for the fundamental (diamond), 1H (x), 2H (square) and 3H (triangle) modes.}
\end{figure}

We have studied the change in the frequencies of 10\Msun\ ZAMS models differentially rotating at 120 \kms\ and 240 \kms.  Overall, the frequencies increase for $l_o$ = 0 and 1, and decrease for $l_o$ = 2 and 3, a trend seen at both 120 \kms\ and 240 \kms.  Our results for $l_o$ =0 are shown in Figure \ref{fig:120fund} for the fundamental, 1H, 2H and 3H modes for a model rotating at 120 \kms.  In this case, the frequency changes are largest for the 3H modes, but are noticeable for all modes by $\beta$ $\approx$ 1.  Similar trends are found for the other values of $l_o$ considered here.  Still, the differences remain relatively small, and it seems unlikely that even extreme differential rotation with this surface rotation velocity will be detectable using the values of the eigenfrequencies alone.

\begin{figure}
\plotone{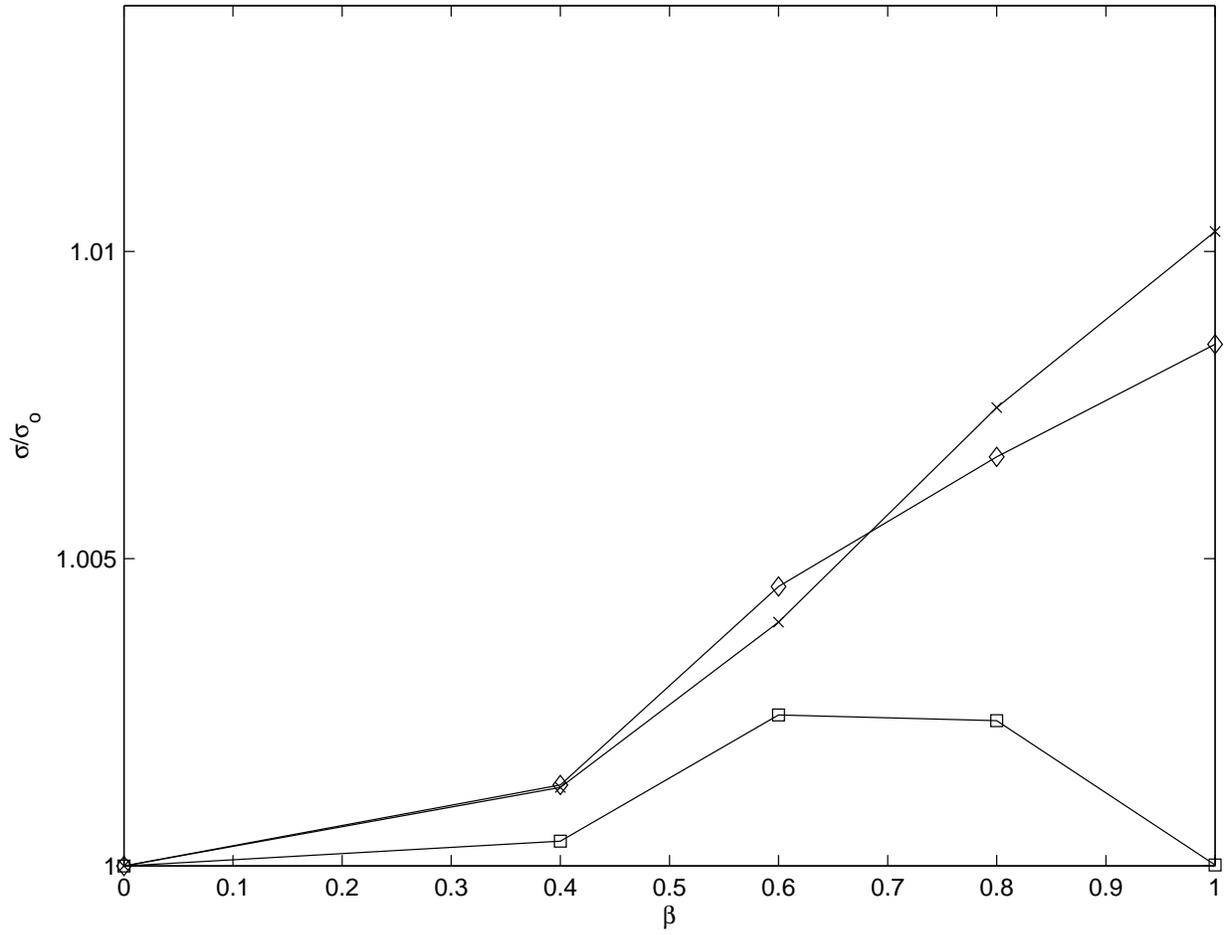}
\caption{\label{fig:240fund}Relative frequencies of the $l_o$ = 0 modes versus differential rotation parameter $\beta$ for a model rotating at 240 \kms.  The curves show the relative frequencies for the fundamental (diamond), 1H (x) and 2H (square) modes.}
\end{figure}

\begin{figure}
\plotone{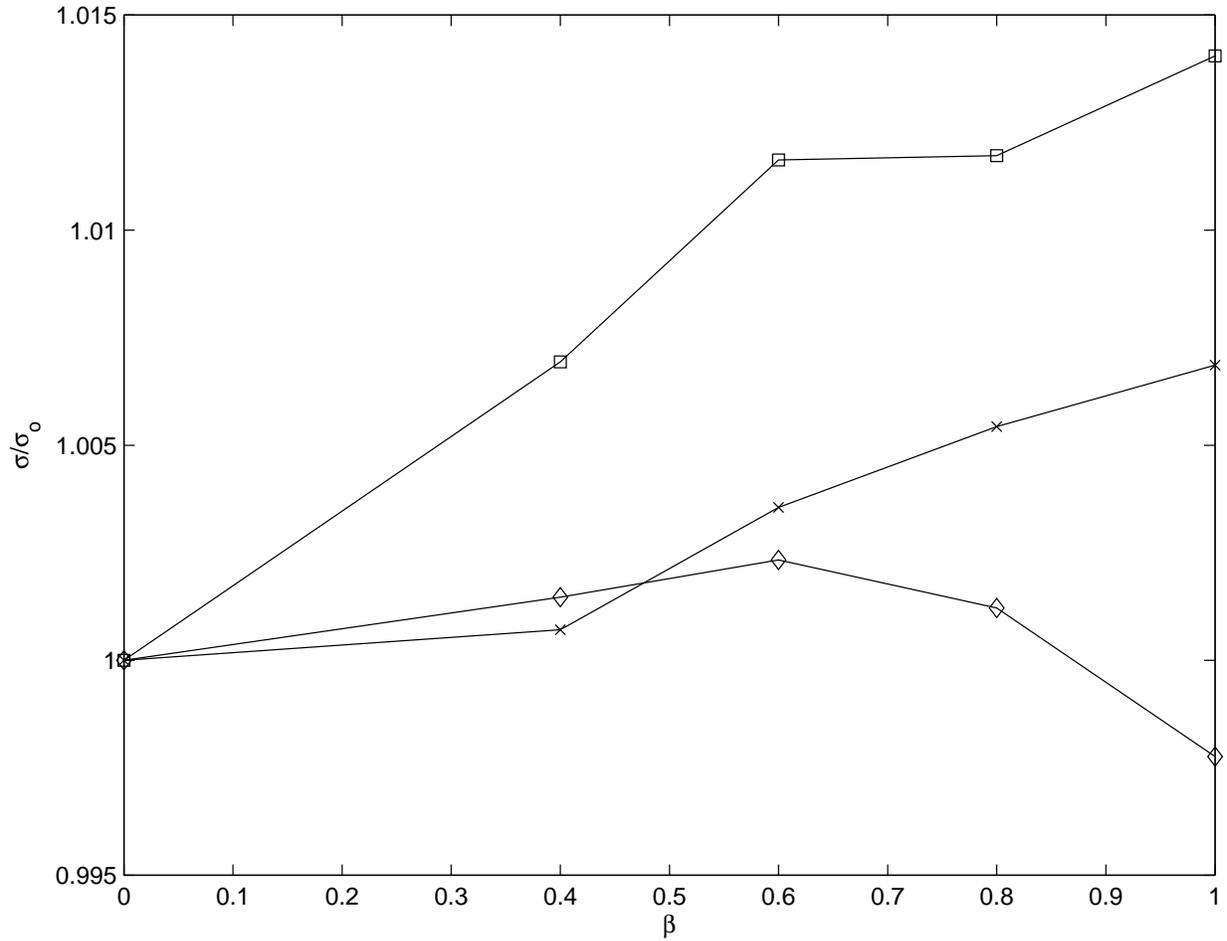}
\caption{\label{fig:l1_240}Relative frequencies of the $l_o$ = 1 modes versus differential rotation parameter for a model rotating at 240 \kms.  Shown are the $p_1$ (diamond), $p_2$ (x) and $p_3$ (square) modes.}
\end{figure}

\begin{figure}
\plotone{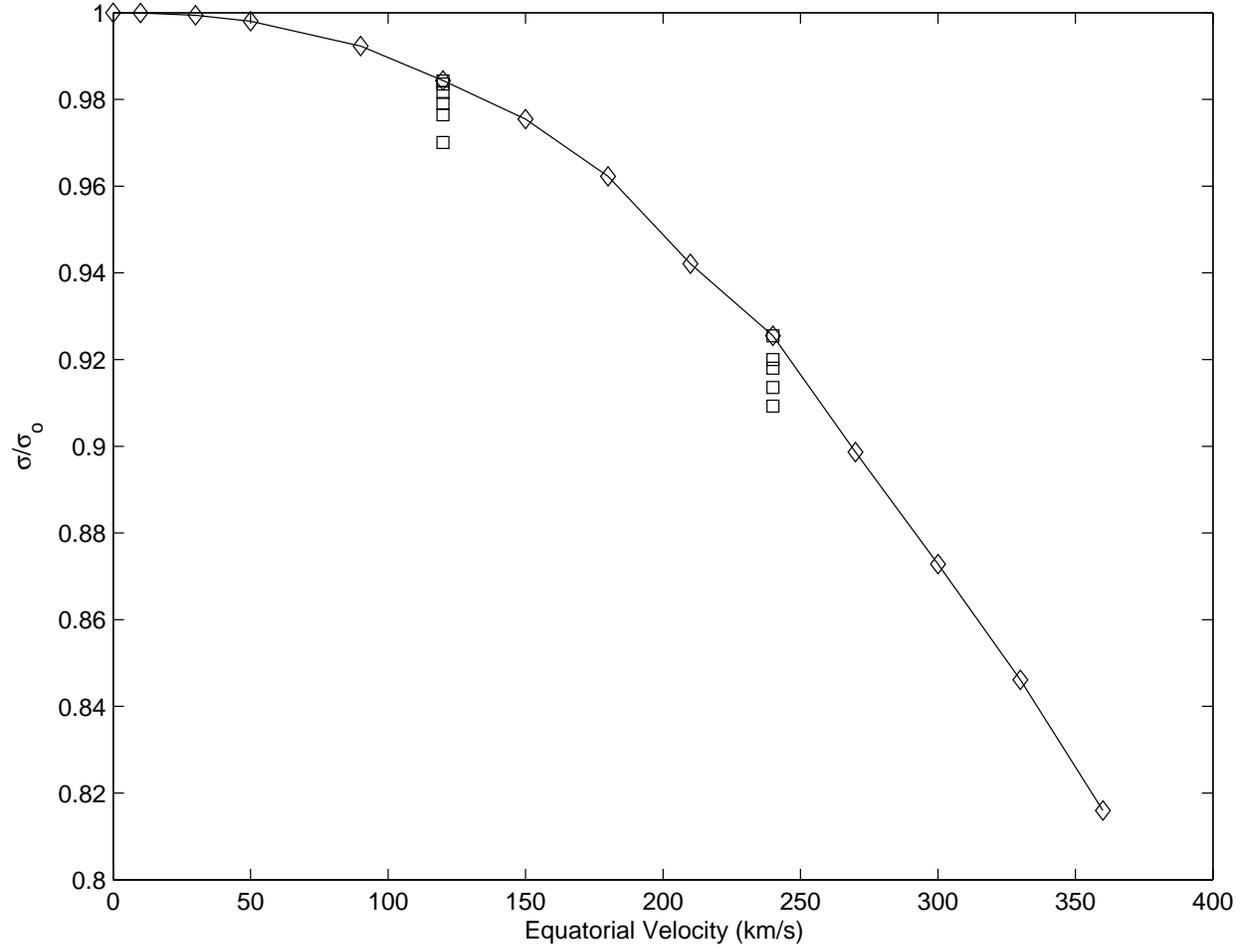}
\caption{\label{fig:comp}  The relative effects of differential rotation for the $l_0$ = 2 $p_2$ mode.  The frequencies for differentially rotating models as a function of $\beta$ at 120 \kms and 240 \kms (squares) are superimposed on the uniformly rotating frequencies (diamonds).}
\end{figure}

The frequency results for the 240 \kms model, shown in Figure \ref{fig:240fund} for $l_o$ = 0, are slightly more promising.  Although we were unable to reliably identify modes above $\beta$ = 1.0, the frequencies already differ by more than 1\% by $\beta$ = 1.0 for the 1H mode, and it seems the differences would be noticeable by $\beta$ = 0.4.  If this trend continues, as seems likely at least for the F and 1H modes, the frequency differences should be large enough to be detectable in these more rapidly rotating stars.  
As noted above, for some modes differential rotation causes the frequencies to increase as in Figures \ref{fig:240fund} and \ref{fig:l1_240}, while for others the frequencies decrease, as in Figure \ref{fig:comp}.  These plots do not include the 3H mode, as we found that the scatter in this mode remained a significant fraction of the variation, despite the improved radial zoning, and so have chosen not to include it in our discussion.  

The effects of differential rotation compared with uniform rotation are shown in Figure \ref{fig:comp} for the $l_o$ = 2 $p_2$ mode.  Differential rotation can change the frequencies by about a percent above and beyond the difference predicted based on surface equatorial velocity alone.  The differences are small; about 1\% for the most extreme differentially rotating model at 120 \kms. Based on the frequencies we have calculated, it may be possible to discriminate between uniform and this type of differential rotation, given the right combination of properly identified frequencies. Since frequencies increase relative to the uniformly rotating case for some $l_o$, and decrease for others, these differences could be used to constrain the rotation.  This would require a star with a few positively identified modes, some of which were either $l_o$ = 0 or 1, and some of which were either $l_o$ = 2 or 3.  The number of modes required and the challenges presented by accurate mode identification in massive main sequence stars may make this extremely difficult in practice.  

\section{Large Separations}
\label{large}

If one considers two stellar models that are in the same evolutionary phase, and appear reasonably close to each other in the HR diagram, the frequencies can be approximately determined from the relevant pulsation constant, Q.  As the mass and radius change with position in the HR diagram, so will the frequencies.  It is expected that for small changes in mass and radius the frequency differences (either frequency separations or ratios) will change, like Q, much more slowly than the individual frequencies.  As a result, the large and small separations are probably more useful than individual frequencies as they are less sensitive to small changes in the models.

\subsection{Uniform Rotation}

\begin{figure}
\plotone{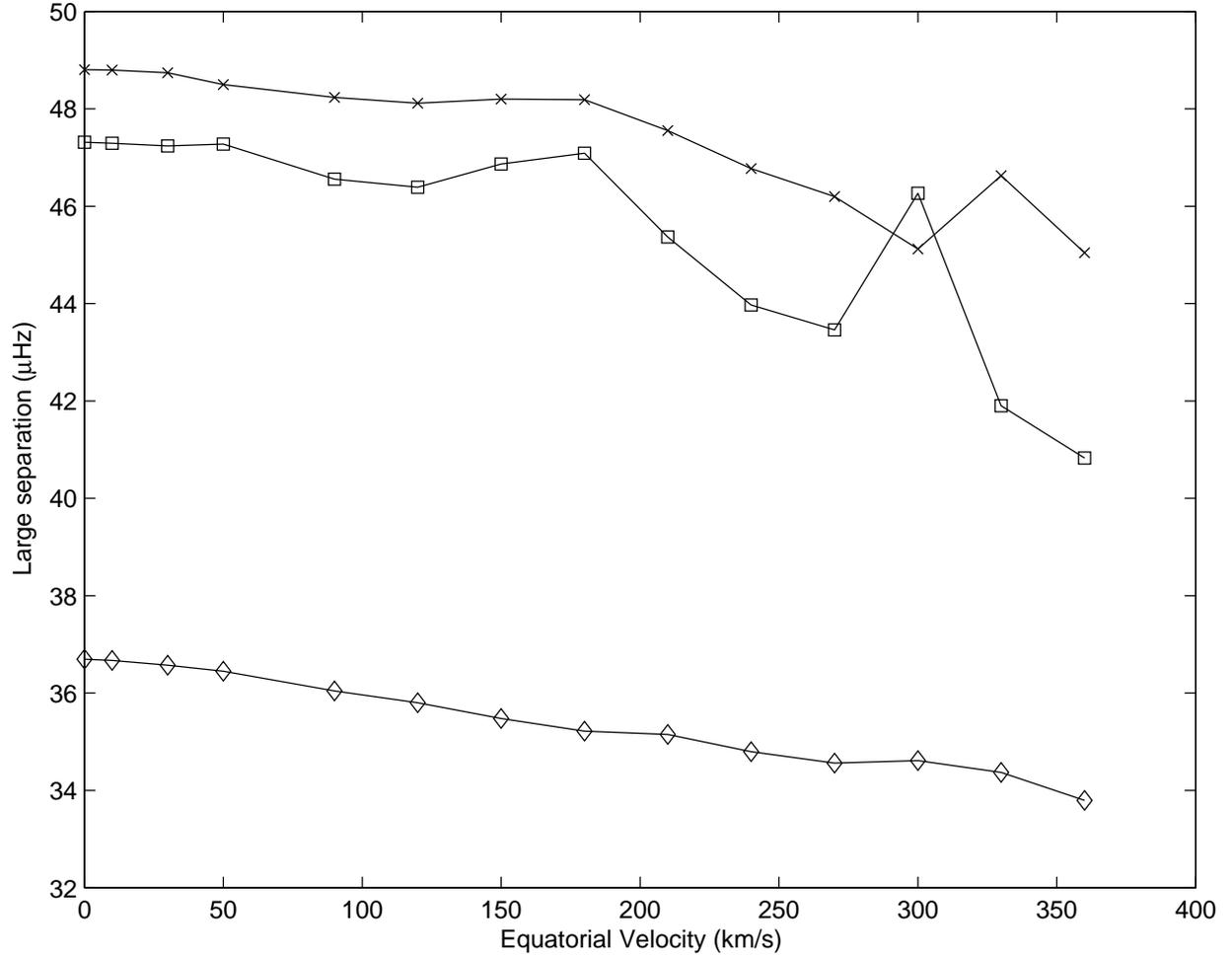}
\caption{\label{fig:largel0}The large separation between the 3H and 2H (square), the 2H and 1H (x), and 1H and F modes (diamonds) for modes with $l_o$ = 0.}
\end{figure}

\begin{figure}
\plotone{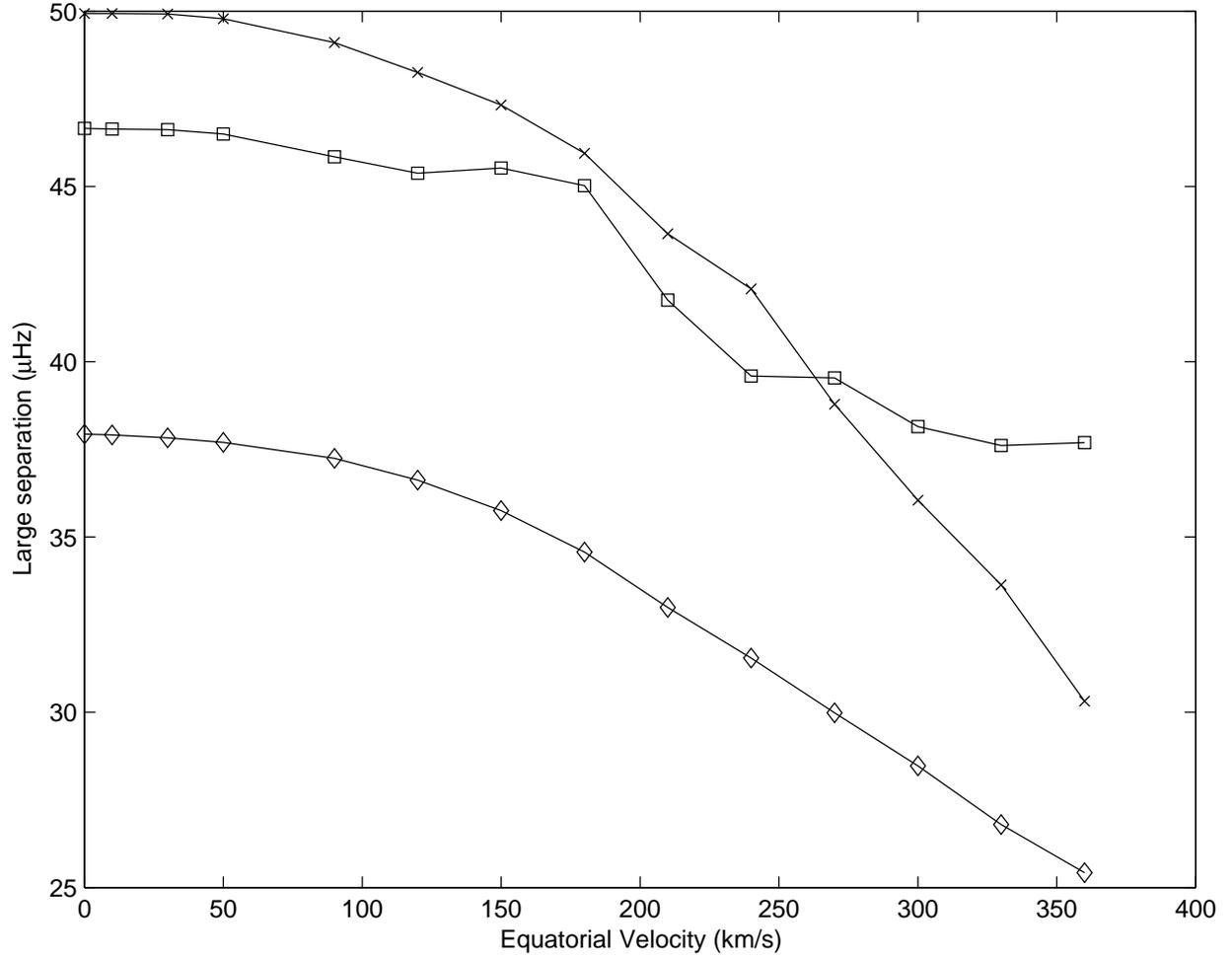}
\caption{\label{fig:largel1}The large separation for the $l_o$ = 2 modes as a function of rotation velocity.  Shown are the separations between the $p_1$ and $f$ modes (diamonds), the $p_2$ and $p_1$ modes (x) and the $p_3$ and $p_2$ modes (squares).}
\end{figure}

The large separation, defined as
\begin{equation}
\Delta\nu_l = \nu_{l,n+1} - \nu_{l,n}
\end{equation}
can provide information about the outer layers of the stellar envelope.  The large separation for the $l_o$ = 0 mode is shown as a function of the rotation rate in Figure \ref{fig:largel0}.  We note that the overall trend of the large separation for this mode is to decrease as the rotation rate increases.  At other $l_o$ the trend is the same and the large separation decreases for every pair of modes considered.  The magnitude of the decrease in large separation does increase slightly with increasing $l_o$, as can be seen by comparing Figures \ref{fig:largel0} and \ref{fig:largel1}.  For nonrotating ZAMS models, both the frequencies and the large separations decrease as the mass (hence the radius, luminosity and effective temperature) increases.  The decrease in the large separation occurs not only because the frequencies decrease, but also because the period ratios increase for increasing ZAMS mass.  However, for stars observed approximately equator on, rotation decreases the perceived luminosity and effective temperature.  This offset between the perceived luminosity and temperature and the large separation may be useful as a rotation discriminant.  Of course, stars observed nearly pole on show an increase in perceived luminosity and effective temperature, which is in line with the decreasing large separation as the rotation rate increases.  Any discriminant of rotation may only be a matter of degree for low inclination objects.  

We noted in section \ref{sec:unifreq} that the decrease in the frequencies with increasing rotation cannot be explained purely by the decrease in the mean density (i.e., a constant Q).  The mean density decreases faster than the pulsation periods increase as the rotation rate increases.  Interestingly, the decrease in the large separation for $l_o$ = 0 in Figure \ref{fig:largel0} is almost entirely offset by the mean density so that $\Delta\nu(\rho_{\odot}/\rho)^{1/2}$ is nearly constant, as discussed by \citet{ulrich} and \citet{reese08}.  The mean density does not offset the steeper decline in the large separation for the $l_o$ = 2 modes shown in Figure \ref{fig:largel1}.  

\subsection{Differential Rotation}

The large separations provide information about the region near the surface of the star, while the frequencies provide information on more global properties.  As discussed above, the separations are less sensitive to small changes in the mass or radius of the star, but since they probe the surface region, may provide information about changes in this region as a result of rotation.  A comparison of Figures \ref{fig:shapeuniform} and \ref{fig:shapediff} shows that the polar radius is significantly more affected by differential rotation than the radius at lower latitudes.  This kind of effect may be detectable using the large separations.  Indeed, based on the results in the previous section, we expect there to be significant changes in the period differences, as we have found that differential rotation can sometimes introduce a significant shift in only one or two of the harmonics.  

\begin{figure}
\plotone{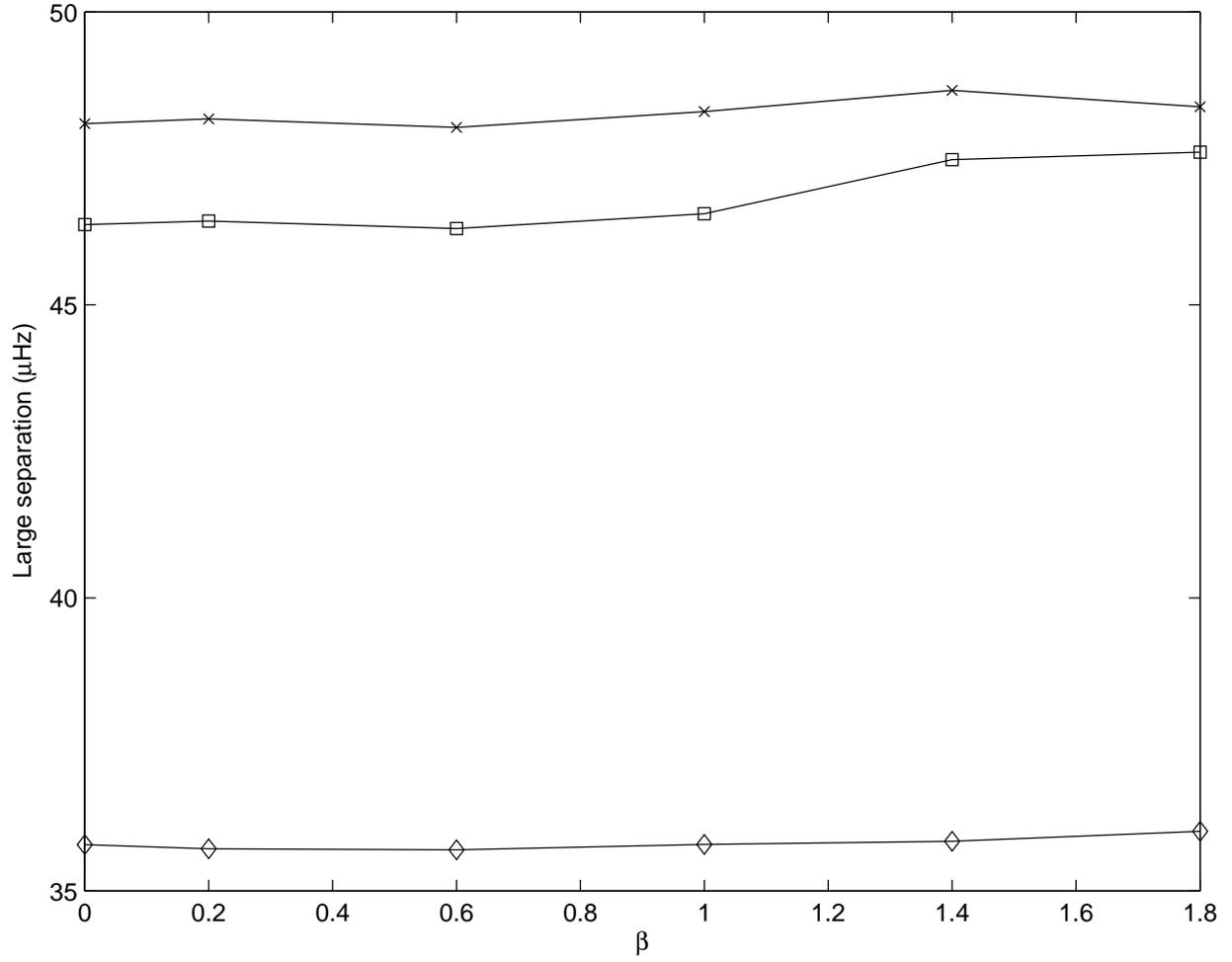}
\caption{\label{fig:large_120}The large separation for the $l_o$ = 0 modes of a differentially rotating model with surface equatorial velocity of 120 \kms, plotted as a function of differential rotation parameter $\beta$.  Symbols are the same as Figure \ref{fig:largel0}.}
\end{figure}

Although the individual frequencies can show significant relative differences, this does not carry over to the large separations.  The large separation for the $l_o$ = 0 modes of a differentially rotating model with surface equatorial velocity of 120 \kms are shown in Figure \ref{fig:large_120}.  The large separations shown in this plot show very little change with increasing $\beta$, much less than the differences shown in Figure \ref{fig:largel0}.  The same lack of variation is seen for all modes.  Given that the large separation probes the surface regions, this might be regarded as somewhat surprising because changing the $\beta$ does change the surface configuration, particularly near the rotation axis.

\begin{figure}
\plotone{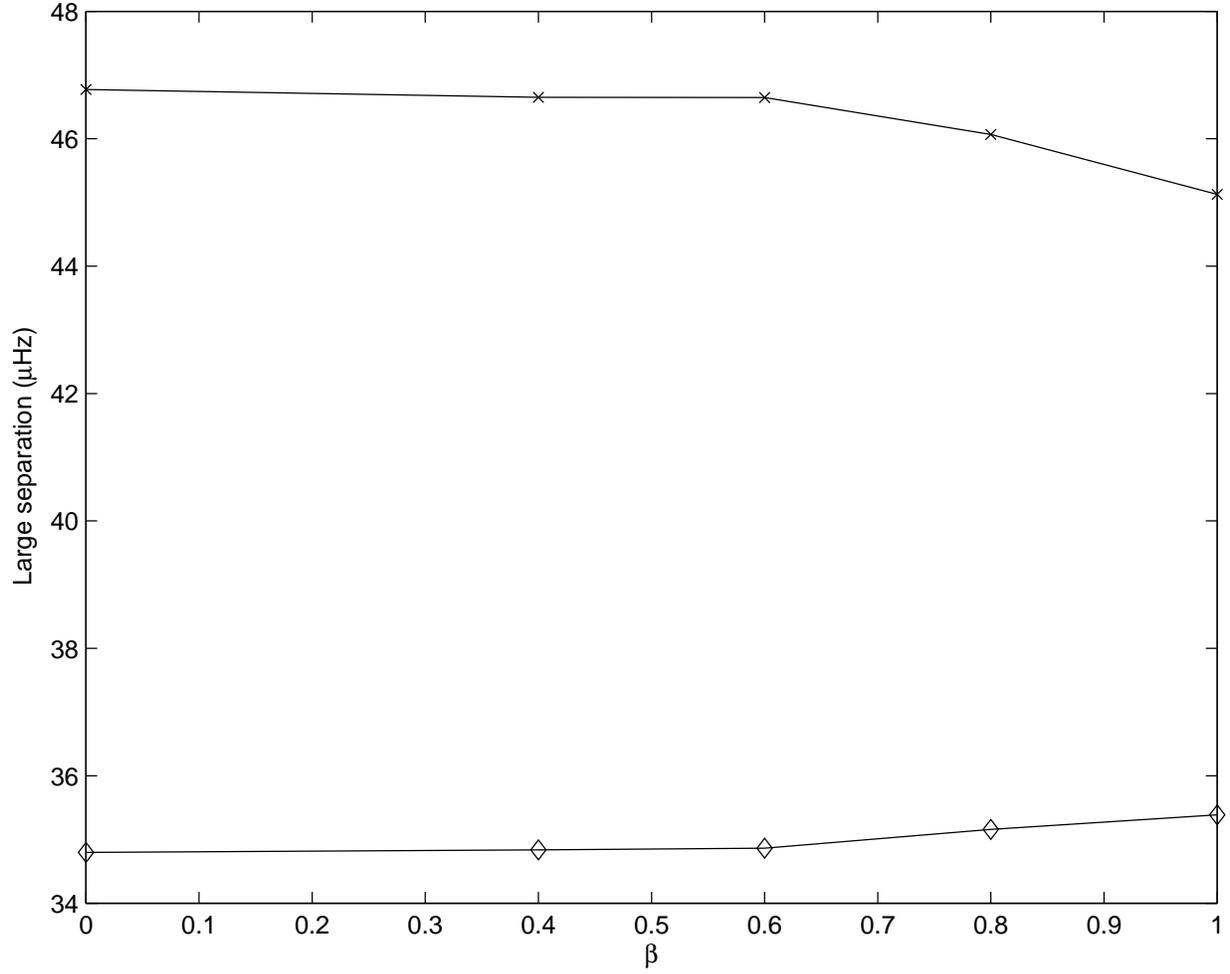}
\caption{\label{fig:large_240}The large separation of the $l_o$ = 0 modes for a differentially rotating model with surface equatorial velocity of 240 \kms.  Symbols are the same as Figure \ref{fig:largel0}.}
\end{figure}

At 240 \kms, the large separations, shown in Figure \ref{fig:large_240}, are again quite constant over the region shown.  Most separations either remain constant or show a slight increase, at least to $\beta$ = 0.6, at which point some of the higher order separations decrease slightly.  Again, this is different from the trend seen in the uniformly rotating models.  Particularly at high $\beta$, the separations moving in different directions may allow constraints to be placed on observed stars.  The differences begin to become noticeable at $\beta$ $\approx$ 0.6 for most of the high order modes considered.  However, for most modes the large separation never differs by more than a few $\mu$Hz.  As for the frequencies, it seems that the large separations are unlikely to produce any very refined constraints on the internal rotation rate, at least for this particular rotation law.

\section{Small Separation}
\label{small}

Asymptotic theory \citep{tassoul80}, which predicts that the large separation should be approximately constant as $n$ gets large, also predicts near degeneracy between modes with the same value of $n + l/2$:
\begin{equation}
\nu_{l,n} \simeq \nu_{l+2,n-1}
\end{equation}
The deviations from this degeneracy are defined as the small separation:
\begin{equation}
\label{eqn:small}
d_{l,n} = \nu_{l,n}-\nu_{l+2,n-1} \simeq -(4l+6)\frac{\Delta\nu}{4\pi^2\nu_{l,n}}\int^R_0\frac{dc}{dr}\frac{dr}{r}
\end{equation}
where $c$ is the sound speed.
The sound speed changes in the core are sufficiently large that the (1/r) variation dominates the integral on the right hand side of Equation \ref{eqn:small}, and hence the small separation is dominated by the structure in the core.  

\subsection{Uniform Rotation}

\begin{figure}
\plotone{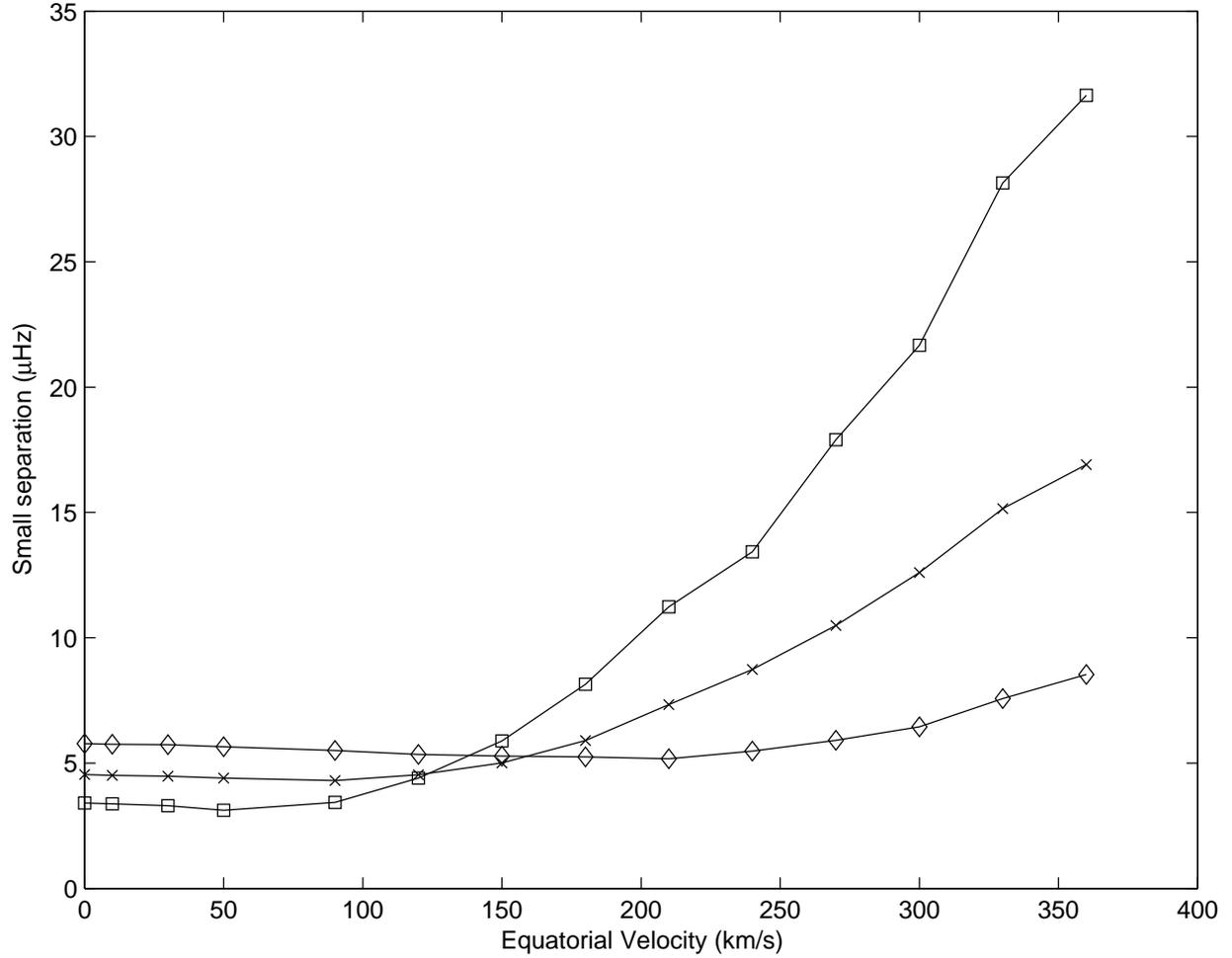}
\caption{\label{fig:small}Small separation for the $l_o$ = 0 and 2 modes as a function of surface equatorial velocity.  Shown are the separations between the $l_o$ = 0, 3H - $l_o$ = 2, $p_2$ modes (squares), $l_o$ = 0, 2H - $l_o$ = 2, $p_1$ modes (x) and $l_o$ = 0, 1H - $l_o$ = 2, $f$ modes (diamonds). }
\end{figure}

At slow uniform rotation, the size and shape of the convective core is nearly unaltered by the rotation, and one would expect the effects on the small separation to be minimal.  Figure \ref{fig:small} shows this to be true, but also shows that the small separation for higher $n$ increases markedly with the rotation rate once the rotation exceeds approximately 150 \kms.  There are slight changes to both the shape and relative size of the convective core with rotation, although the absolute mass and radius of the core change only slightly.  It is not obvious why the small separation increases so markedly.  This effect has also been noted by \citet{lignieres}

Small separations are frequently used as probes of the core structure of stars, and can be used to constrain overshooting and core composition \citep{soriano}.  Their results indicate that convective core overshooting causes a slight decrease in the small separations, with the effect becoming more pronounced as the star evolves.  This slight trend is opposite to that produced by at least moderate rotation, which appreciably increases the small separation.  Clearly, this is a situation in which caution must be exercised when using observed modes to constrain conditions deep in the stellar interior.

\subsection{Differential Rotation}

\begin{figure}
\plotone{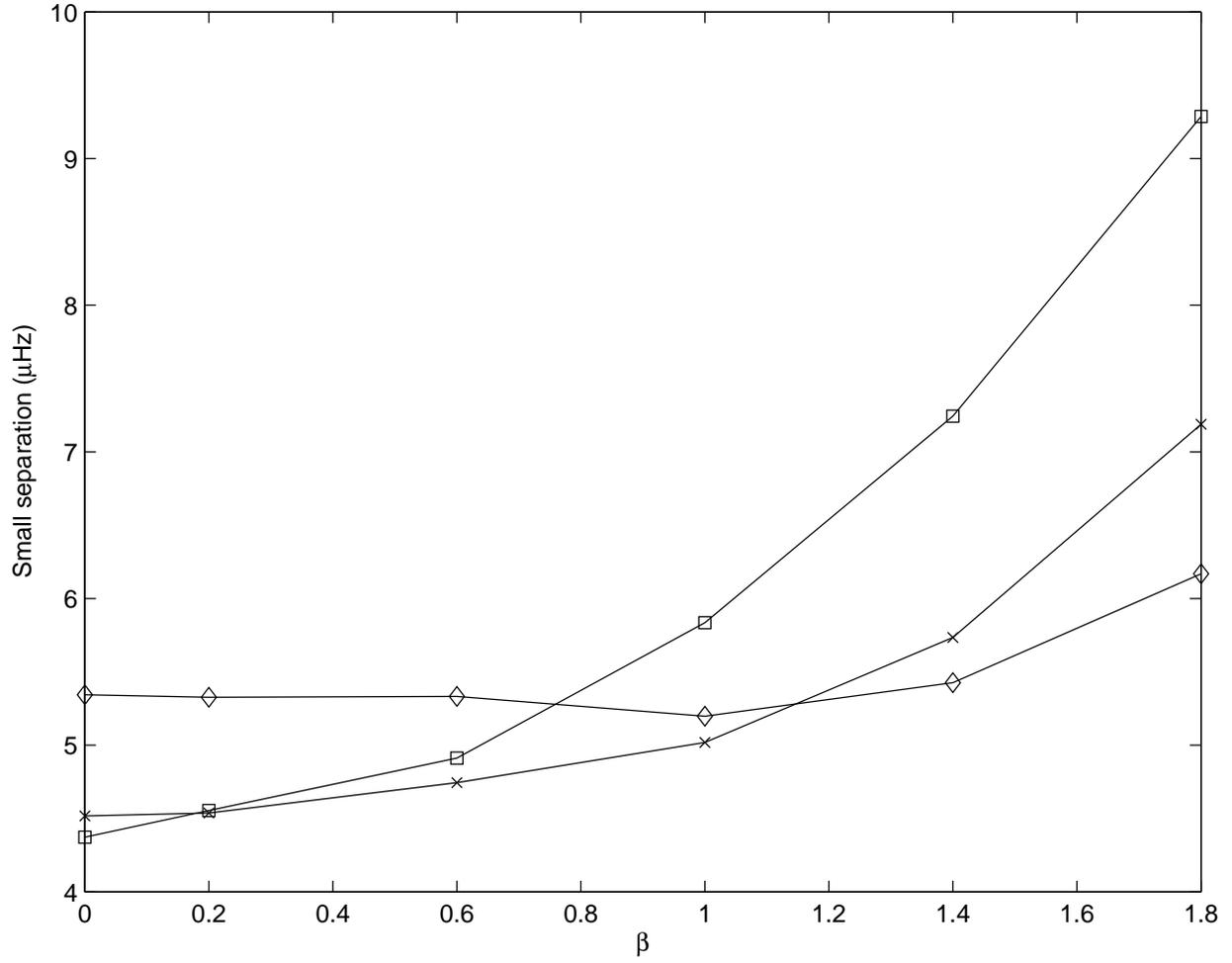}
\caption{\label{fig:small_120}Small separations for the even modes for a differentially rotating model with surface equatorial velocity of 120 \kms, plotted as a function of differential rotation parameter $\beta$.  Symbols are defined as in Figure \ref{fig:small}.}
\end{figure}

\begin{figure}
\plotone{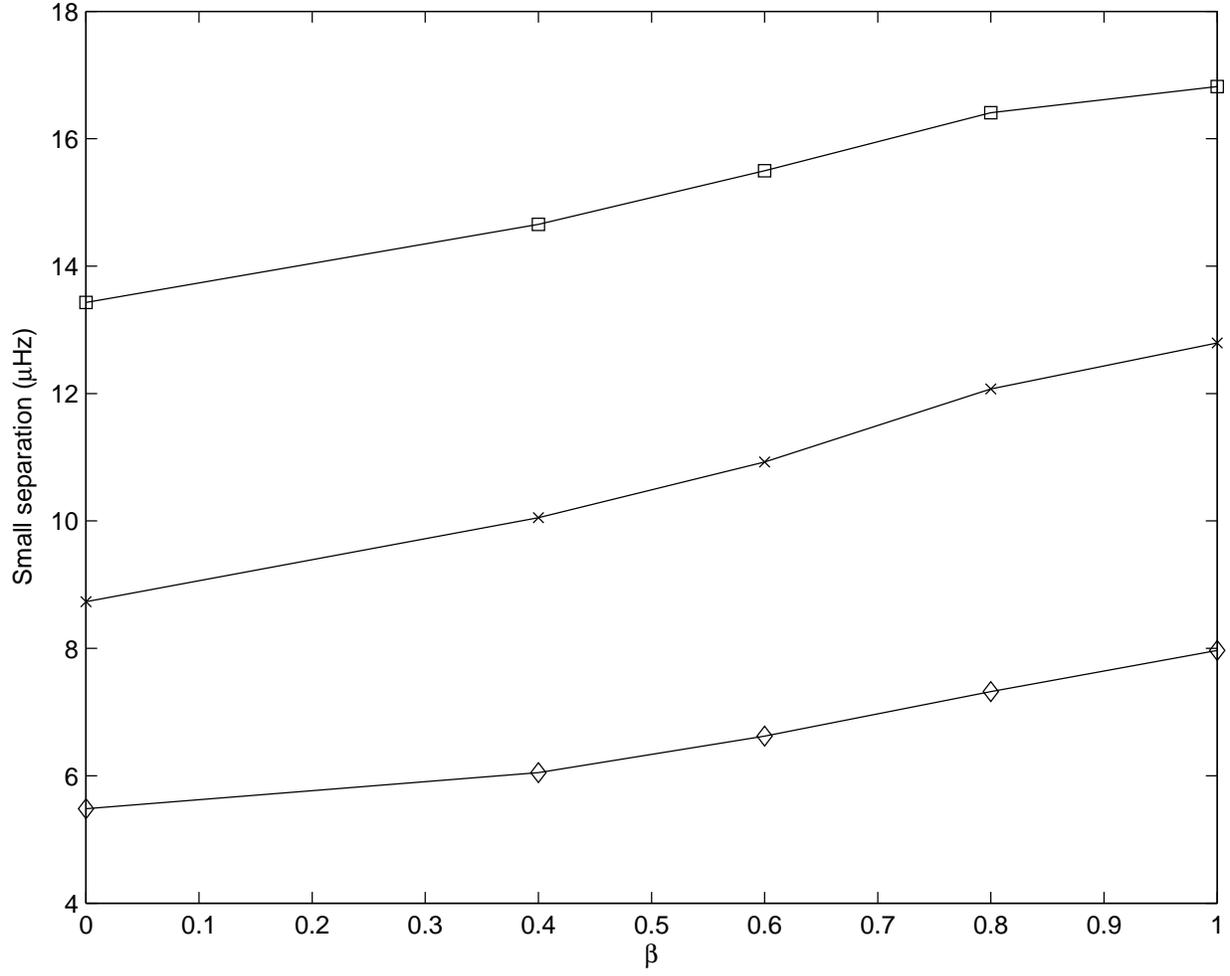}
\caption{\label{fig:small_240}Small separations for the even modes of a differentially rotating model with surface equatorial velocity of 240 \kms.  Symbols are defined as in Figure \ref{fig:small}.}
\end{figure}

For differentially rotating models, the overall trend is the same, with small separations increasing with increasing differential rotation.  However, Figures \ref{fig:small_120} and \ref{fig:small_240} show that the variation in small separation is much less than for uniformly rotating models.  The trends are consistent with the relationship between the effects of $\beta$ and those of increasing the uniform rotation rate as shown in Figure \ref{fig:comp}.  The effects of increasing $\beta$ on the convective core mimic to some extent those of increasing the uniform rotation rate, although high values of $\beta$ do make the convective core more oblate.  The different effects of $\beta$ on the large and small separations is understandable in that increasing $\beta$ increases the rotation and its effects near the rotation axis, and this certainly includes the convective core.  However, this trend of increasing $\beta$ producing similar trends to increasing uniform rotation rate does not give us confidence that we have a useful tool for diagnosing a rotation law of the kind we have considered through the small separation.

\section{Conclusion}
\label{conclusion}

We have investigated the effects of uniform and differential rotation on pulsational eigenfrequencies.  For uniformly rotating models, we have found that the frequencies decrease as the rotation rate increases for all values of $l_o$ and $n$ considered here, although the rate of decrease varies with the mode in question.  While this frequency behavior is expected assuming the period-mean density relation applies, the frequency changes are much smaller than the period-mean density would suggest.  We do find a pulsation constant being approximately constant if we use the surface radius at a colatitude of about 40$^{\circ}$ in the period-mean density relation, although this radius does not represent the mean density.  

For the differential rotation law considered here, we find the frequencies at a given velocity may either increase or decrease, depending on $l_o$, with increasing differential rotation, relative to the uniformly rotating model.  However, the overall effects in all cases are comparatively small, with maximum differences typically on the order of 1\% when compared to the uniformly rotating case.

Uniform rotation decreases the large separation by several $\mu$Hz ($<$ 10) over the entire range of rotation (0 - 360 \kms) considered here.  The large separation was virtually unchanged ($<$ 1 $\mu$Hz) from that of uniform rotation for the range of differential rotation parameters considered here, despite the noticeable change in the surface shape.  Although this change in shape is noticeable, it is still considerably smaller than the change produced by uniform rotation.  Both uniform and differential rotation increase the small separation.  The small separation can change markedly over the range of uniform rotation considered, while the dependence of the small separation on the rotation profile is more modest but not inconsistent with the other effects produced when comparing uniform and differential rotation.  The effects of rotation on the frequencies and separations are generally large enough that rotation must be considered in the asteroseismology of these upper main sequence stars.  While the precise rotation rate at which one must be concerned with rotation depends on the level of accuracy achievable, it is certainly no larger than 100 \kms\ for our 10 \Msun\ ZAMS models.

Although we have shown there can be significant differences in the pulsation properties of rotating stars, it is not clear that these results can actually be used to constrain the interior rotation rate.  Given the possible combinations of effects from the rotation rate and distribution, the mass of the star, convective overshoot, evolutionary stage, etc, it seems unlikely that pulsation properties will give a unique solution, particularly if the number of observed modes is modest or cannot be properly identified.  However, we have found that some combinations of modes constrain some of these properties.

\acknowledgements

This research was supported by the National Science and Engineering Research Council of Canada through a Discovery grant and a graduate fellowship.  Computational facilities were provided with grants from the Canadian Foundation for Innovation and the Nova Scotia Innovation Research Trust.  

\end{document}